\documentclass[aps,twocolumn,floatfix,amsmath,amsfonts]{revtex4-1}

\usepackage{graphicx}
\usepackage{bm}
\usepackage{units}
\usepackage{upgreek}
\usepackage[normalem]{ulem}
\begin{document}

\title{Magneto-Stark-effect of yellow excitons in cuprous oxide}
\author{Patric Rommel}
\author{Frank Schweiner}
\thanks{Present address: MINT-Kolleg, Universit\"at Stuttgart}
\author{J\"org Main}
\affiliation{Institut f\"ur Theoretische Physik 1, Universit\"at Stuttgart, 70550 Stuttgart, Germany}
\author{Julian Heck\"otter}
\author{Marcel Freitag}
\author{Dietmar Fr\"ohlich}
\author{Kevin Lehninger}
\author{Marc A{\ss}mann}
\author{Manfred Bayer}
\affiliation{Experimentelle Physik 2, Technische Universit\"at Dortmund, 44221 Dortmund, Germany}
\date{\today}

\begin{abstract}
We investigate and compare experimental and numerical excitonic spectra of the yellow series in cuprous oxide Cu$_2$O
in the Voigt configuration and thus partially extend the results from Schweiner \emph{et al.}\ [Phys. Rev. B \textbf{95}, 035202 (2017)], who only considered the Faraday configuration. The main difference between the configurations is given by an additional
effective electric field in the Voigt configuration, caused by the motion of the exciton through
the magnetic field. This Magneto-Stark effect was already postulated by Gross \emph{et al.}\ and Thomas \emph{et al.}\ in 1961 [Sov.\ Phys.\ Solid State \textbf{3}, 221 (1961); Phys.\ Rev.\ \textbf{124}, 657 (1961)]. Group theoretical
considerations show that the field most of all significantly increases the number of allowed lines by decreasing the symmetry of the system.
This conclusion is supported by both the experimental and numerical data.\\
\end{abstract}

\pacs{71.35.Ji, 71.35.-y, 78.40.-q, 02.20.-a}

\maketitle

\section{Introduction}

Application of electric or magnetic fields, both representing well-controlled external perturbations, has offered detailed insight into quantum mechanical
systems by inducing characteristic shifts and lifting degeneracies of energy levels. This potential became apparent already for the simplest quantum mechanical
system in nature, the hydrogen atom, for which an external field reduces the
symmetry from spherical to cylindrical \cite{Fri89,Has89,ruder1994atoms}.

In condensed matter, the hydrogen model is often applied as a simple, but nevertheless successful model for problems in which Coulomb interaction mediates the
coupling between opposite charges of quasi-particles \cite{WannierExcitonen1937,knox1963excitons}. The most prominent example is the exciton, the bound complex of a negatively charged electron and a positively
charged hole.  In particular, for crystals of high symmetry such as cubic systems, the phenomenology of energy levels in an external field often resembles the corresponding
hydrogen spectra after rescaling the energy axis with the corresponding material parameters that enter the Rydberg energy such as the effective mass and the
dielectric constant which in these systems are isotropic.

Recent high resolution spectroscopy of excited exciton states in Cu$_2$O has allowed one to reveal features which result from the crystal environment
having discrete instead of continuous symmetry, leading to deviations from a simple hydrogen description \cite{GiantRydbergExcitons,ObservationHighAngularMomentumExcitons,Heckoetter2017}.
Theoretical investigations of excitons in external fields \cite{Sylwia2016,Sylwia2017,Kurz2017} also show differences to
a simple hydrogen model.
For example, the level degeneracy is lifted
already at zero field. A detailed analysis shows that this splitting arises mostly from the complex valence band structure
\cite{SchoeneLuttinger,frankmagnetoexcitonscuprousoxide,ImpactValence,Bauer1988}. In effect, angular momentum
is not a good quantum number anymore, but states of different angular momenta having the same parity are mixed. As the exciton size is large compared to
that of the crystal unit cell, the mixing is weak, however, so that angular momentum may still be used  approximately as quantum number. Simply speaking,
this can be understood in the following way: The exciton wave function “averages” over the crystal lattice, assembled by the cubic unit cells. With increasing
wave function extension a state becomes less and less sensitive to the arrangement of the atoms in a crystal unit cell and thus to the deviations from spherical symmetry.

For example, the size of the cubic unit cell in Cu$_2$O is about 0.4 nm \cite{korzhavyi2011literature}, while the Bohr radius is {$a_{\mathrm{exc}} =$} $ 1.1\,\mathrm{nm}$ for the dipole-active $P$-type excitons dominating the absorption \cite{KavoulakisBohrRadius}.
{Using the formula $\langle r \rangle_{n,L} = \frac{1}{2} a_{\mathrm{exc}} (3 n^2 - L(L+1))$ \cite{GiantRydbergExcitons},} the size of the lowest exciton $n=4$, for which the mixing becomes optically accessible,
is {$\langle r\rangle_{4,1}  = 25.4\,\mathrm{nm}$}, covering thousands of unit cells. As a consequence of the weak mixing,
the state splitting is small compared to the separation between levels of different $n$. Generally, the average radius of the wavefunction increases as $n^2$, so that the
zero-field state splitting decreases with increasing principal quantum number, leading to the levels becoming quasi-degenerate again for $n$ larger than about 10, as demonstrated
in Ref.~\cite{ObservationHighAngularMomentumExcitons}. For smaller principal quantum numbers, the splitting can be well resolved, in particular, because the mixing of levels also causes a redistribution of oscillator strength,
so that not only the $P$-states but also higher lying odd states such as $F, H, \ldots$ within a multiplett can be resolved in single photon absorption.

Applying a magnetic field leads to a squeezing of the wave function mostly normal to the field, so that the influence of the crystal is enhanced. Even when assuming spatial isotropy,
the symmetry is reduced to rotational invariance about the field, leaving only the magnetic quantum number $m$ as conserved quantity. As a consequence, states of the same $m$ but different
orbital angular momentum become mixed, activating further lines optically. In combination with the Zeeman splitting of the levels this leads to a rich appearance of the absorption spectra, in
particular, as for excitons due to their renormalized Rydberg energy being much smaller than in the atomic case, also resonances between states of different principal quantum numbers can be
induced by the field application, at least for excited excitons.

As indicated above, a key ingredient for the renormalized Rydberg energy are the smaller masses of the involved particles forming an exciton, in particular of the hole compared to the nucleus.
In Cu$_2$O, the electron mass is almost that of the free electron ($m_{\mathrm{e}} = 0.99 m_0$ \cite{HodbyEffectiveMasses}), while the hole represents the lighter particle ($m_{\mathrm{h}} = 0.58 m_0$ \cite{HodbyEffectiveMasses}), in contrast to most other semiconductors.
This makes the reduced mass entering the relative motion of electron and hole only a factor of less than 3 smaller than in hydrogen. The mass of the exciton center of mass motion, on the other hand, is clearly more
than three orders of magnitude smaller. This raises the question whether this mass {disparity} causes a difference of the optical spectra in magnetic field, as the excitons are generated
with a finite wavevector identical to the wavevector of the exciting laser.

In that respect two different field configurations need to be distinguished, namely the Faraday and the Voigt configuration. In the first case the magnetic field is applied along the optical axis,
while in the second case the field is oriented normal to the optical axis.
{One would not expect a difference in the spectra between the two cases when probing atomic systems using a fixed polarization. }

Here we have performed corresponding experiments which demonstrate that in contrast to the atomic case the exciton spectra differ significantly for the two field configurations.
These findings are in good agreement with detailed theoretical
calculations. From the comparison we trace the difference to the Magneto-Stark-effect \cite{Gross61,Hopfield61,Thomas61} which is
acting only in the Voigt configuration,
where the excited exciton is moving normal to the magnetic field, so that its two constituents are subject to the Lorentz force acting in opposite direction for electron and hole and therefore trying to move them apart,
similar to the action of an electric field of the form
\begin{eqnarray}
\mathcal{F}_{\mathrm{MSE}} = \frac{\hbar}{M} \left( \boldsymbol{K} \times \boldsymbol{B} \right),
\label{eq:MotionalStarkField}
\end{eqnarray}
where $M$ is the exciton center-of-mass, $\boldsymbol{K}$ is the
exciton wavevector, and $\boldsymbol{B}$ is the magnetic field. The
action of this field is obviously absent in the Faraday configuration.

{The main novelty in this work comes from the comparison of the spectra for atoms and for excitons. One would not expect any difference in the spectra for atoms when the magnetic field
is aligned parallel or normal to the magnetic field. For excitons, on the other hand, it does make a difference as we demonstrate here. Further, we identify the origin of this difference as the Magneto-Stark-Effect (MSE),
which does not show up prominently for atoms due to their heavy mass and small size. The MSE has been introduced theoretically already in the 1960s (Refs.~\cite{Gross61,Hopfield61,Thomas61,Zhilich1969})
but clear demonstrations of its effect are still scarce \cite{Lafrentz2013}.

The symmetry reduction by a longitudinal field (electric or magnetic) was demonstrated also in previous work. Still, in atomic physics, but also in semiconductor physics,
it is common believe that for a bulk crystal there should not be a difference between the two configurations as the symmetry reduction by the field is the same, considering only the impact of the magnetic field.
The additional symmetry breaking here comes from the optical excitation, which in Faraday-configuration is parallel to the field, leading to no further symmetry reduction (as studied previously), while in Voigt the optical axis is normal to it.
This lifts virtually all symmetries, as evidenced by the observation of basically all possible optical transitions which therefore represents also a point of novelty.

In this respect we want to emphasize that the difference between the two field configurations that has been reported for confined semiconductor quantum structures has a different origin, as the geometric confinement breaks the spatial isotropy
already at zero field. When applying a field, the energy spectrum of the free particles depends strongly on the field orientation \cite{landwehr1991landau}: For example, in a quantum well application of the field normal to the quantum well leads 
to a full discretization of the energy levels \cite{Maan1984,Bauer1988,Bauer1988Magnetoexcitons}, while application in the well plane leaves the carrier motion along the field free (see Ref.~\cite{Bayer1995} and further references therein).
Here one has a competition between the magnetic and the geometric confinement normal to the field, while for normal orientation there is only the magnetic confinement normal to the field. This causes an intrinsic difference of the magneto-optical
spectra in terms of state energies. The same is true to other quantum structures like quantum dots for which the geometric confinement in most cases is spatially anisotropic, while for the bulk the energy spectrum is independent of the field
orientation \cite{Wang1996}.
We also note that in confined semiconductor quantum structures like quantum wells the
difference in the spectra of the Faraday and the Voigt configuration
depends on the difference in the number of degrees of freedoms involved in the interaction between the
excitons and the magnetic field, i.e., two for Faraday and one for Voigt configuration \cite{Czajkowski2003}.

However, also for quantum structures one can indeed use the different field orientations to vary the number of observed optical transitions, somewhat similar to the case studied here. For example, in quantum wells or flat quantum dots
the field orientation normal to the structure leaves the rotational in-plane symmetry about the field unchanged, while this symmetry is broken for application in the structural plane, so that one can mix excitons of different angular moments
and make dark excitons visible \cite{Bayer2000}. }

\section{Hamiltonian \label{sec:Hamiltonian}}
Our theoretical description of excitonic spectra in $\mathrm{Cu_{2}O}$ with an external magnetic field builds upon
Schweiner \emph{et al.}'s treatment in Ref. \cite{frankmagnetoexcitonscuprousoxide}, where only the Faraday configuration was considered. The Hamiltonian without magnetic field is given by
\begin{eqnarray}
H  =  E_{\mathrm{g}} +  H_{\mathrm{e}}\left(\boldsymbol{p}_{\mathrm{e}}\right)+H_{\mathrm{h}}\left(\boldsymbol{p}_h\right) + V\left(\boldsymbol{r}_{\mathrm{e}} - \boldsymbol{r}_{\mathrm{h}}\right),
 \label{eq:Hamiltonian}
\end{eqnarray}
where $H_{\mathrm{e}}$ and $H_{\mathrm{h}}$ are the kinetic energies of the electron and hole, respectively. They are given by
\begin{equation}
H_{\mathrm{e}}\left(\boldsymbol{p}_{\mathrm{e}}\right)=\frac{\boldsymbol{p}_{\mathrm{e}}^{2}}{2m_{\mathrm{e}}}
\label{eq:ElectronKinetic}
\end{equation}
and
\begin{eqnarray}
H_{\mathrm{h}}\left(\boldsymbol{p}_{\mathrm{h}}\right) & = & H_{\mathrm{so}}+\left(1/2\hbar^{2}m_{0}\right)\left\{ \hbar^{2}\left(\gamma_{1}+4\gamma_{2}\right)\boldsymbol{p}_{\mathrm{h}}^{2}\right.\phantom{\frac{1}{1}}\nonumber \\
 & + & 2\left(\eta_{1}+2\eta_{2}\right)\boldsymbol{p}_{\mathrm{h}}^{2}\left(\boldsymbol{I}\cdot\boldsymbol{S}_{\mathrm{h}}\right)\phantom{\frac{1}{1}}\nonumber \\
 & - & 6\gamma_{2}\left(p_{\mathrm{h}1}^{2}\boldsymbol{I}_{1}^{2}+\mathrm{c.p.}\right)-12\eta_{2}\left(p_{\mathrm{h}1}^{2}\boldsymbol{I}_{1}\boldsymbol{S}_{\mathrm{h}1}+\mathrm{c.p.}\right)\phantom{\frac{1}{1}}\nonumber \\
 & - & 12\gamma_{3}\left(\left\{ p_{\mathrm{h}1},p_{\mathrm{h}2}\right\} \left\{ \boldsymbol{I}_{1},\boldsymbol{I}_{2}\right\} +\mathrm{c.p.}\right)\phantom{\frac{1}{1}}\nonumber \\
 & - & \left.12\eta_{3}\left(\left\{ p_{\mathrm{h}1},p_{\mathrm{h}2}\right\} \left(\boldsymbol{I}_{1}\boldsymbol{S}_{\mathrm{h}2}+\boldsymbol{I}_{2}\boldsymbol{S}_{\mathrm{h}1}\right)+\mathrm{c.p.}\right)\right\} \phantom{\frac{1}{1}}
 \label{eq:HoleKinetic}
\end{eqnarray}
with the spin-orbit interaction
\begin{equation}
H_{\mathrm{so}}=\frac{2}{3}\Delta\left(1+\frac{1}{\hbar^{2}}\boldsymbol{I}\cdot\boldsymbol{S}_{\mathrm{h}}\right).
\end{equation}
Here, $\boldsymbol{I}$ is the quasi-spin and $\boldsymbol{S}_{\mathrm{h}}$ the spin $S_{\mathrm{h}}=\frac{1}{2}$ of the hole and c.p.\ denotes cyclic permutation.
Electron and hole interact via the screened Coulomb potential
\begin{equation}
V\left(\boldsymbol{r}_{\mathrm{e}} - \boldsymbol{r}_{\mathrm{h}}\right) = 
-\frac{e^2}{4\pi\varepsilon_0 \varepsilon \left|\boldsymbol{r}_{\mathrm{e}} - \boldsymbol{r}_{\mathrm{h}} \right|} ,
\end{equation}
with the dielectric constant $\varepsilon$.
To account for the magnetic field $\boldsymbol{B}$, we use the minimal substitution $\boldsymbol{p}_{\mathrm{e}} \rightarrow \boldsymbol{p_{\mathrm{e}}} + e \boldsymbol{A}(\boldsymbol{r}_e)$ and
$\boldsymbol{p}_{\mathrm{h}} \rightarrow \boldsymbol{p}_{\mathrm{h}} - e \boldsymbol{A}(\boldsymbol{r}_h)$ with the vector potential for a homogenous field $\boldsymbol{A}(\boldsymbol{r}_{\mathrm{e,h}}) = \left(\boldsymbol{B} \times \boldsymbol{r}_{\mathrm{e,h}} \right)/2$.
The energy gained by the electron and hole spin in the external magnetic field is described by
\begin{equation}
H_{B}=\mu_{\mathrm{B}}\left[g_{c}\boldsymbol{S}_{\mathrm{e}}+\left(3\kappa+g_{s}/2\right)\boldsymbol{I}-g_{s}\boldsymbol{S}_{\mathrm{h}}\right]\cdot\boldsymbol{B}/\hbar.
\end{equation}
with the Bohr magneton $\upmu_B$ and the g-factor of the hole spin $g_s \approx 2$.
We finally switch into the center of mass reference frame \cite{Schmelcher1992}:
\begin{align}
 \boldsymbol{r} &= \boldsymbol{r}_{\mathrm{e}} - \boldsymbol{r}_{\mathrm{h}},\nonumber\\
 \boldsymbol{R} &= \frac{m_{\mathrm{e}}}{m_{\mathrm{e}} + m_{\mathrm{h}}} \boldsymbol{r}_{\mathrm{e}} + \frac{m_{\mathrm{h}}}{m_{\mathrm{e}} + m_{\mathrm{h}}} \boldsymbol{r}_{\mathrm{h}},\nonumber\\
 \boldsymbol{p} &= \hbar \boldsymbol{k} - \frac{e}{2} \boldsymbol{B} \times \boldsymbol{R} = \frac{m_h}{m_{\mathrm{e}} + m_{\mathrm{h}}} \boldsymbol{p}_{\mathrm{e}} - \frac{m_{\mathrm{e}}}{m_{\mathrm{e}} + m_{\mathrm{h}}} \boldsymbol{p}_{\mathrm{h}}, \nonumber\\
 \boldsymbol{P} &= \hbar \boldsymbol{K} + \frac{e}{2} \boldsymbol{B} \times \boldsymbol{r} = \boldsymbol{p}_{\mathrm{e}} + \boldsymbol{p}_{\mathrm{h}},
 \label{eq:CentorOfMass}
\end{align}
and set $\boldsymbol{R} = 0$.
More details can be found in Refs.~\cite{frankmagnetoexcitonscuprousoxide, franklinewidth, frankevenexcitonseries, Luttinger52CyclotronResonanceSemiconductors}
and values of material parameters for Cu$_2$O used in Eqs.~\eqref{eq:Hamiltonian}-\eqref{eq:CentorOfMass} are listed in Table~\ref{tab:Constants}.

\begin{table}

\protect\caption{Material parameters used in Eqs.~\eqref{eq:Hamiltonian}-\eqref{eq:CentorOfMass}. }

\begin{centering}
\begin{tabular}{lll}
\hline 
band gap energy & $E_{\mathrm{g}}=2.17208\,\mathrm{eV}$ & \cite{GiantRydbergExcitons}\tabularnewline
electron mass & $m_{\mathrm{e}}=0.99\, m_{0}$ & \cite{HodbyEffectiveMasses}\tabularnewline
hole mass & $m_{\mathrm{h}} = 0.58\, m_0$ & \cite{HodbyEffectiveMasses}\tabularnewline
dielectric constant & $\varepsilon=7.5$ & \cite{LandoltBornstein1998DielectricConstant}\tabularnewline
spin-orbit coupling & $\Delta=0.131\,\mathrm{eV}$ & \cite{SchoeneLuttinger}\tabularnewline
valence band parameters & $\gamma_{1}=1.76$ & \cite{SchoeneLuttinger}\tabularnewline
 & $\gamma_{2}=0.7532$ & \cite{SchoeneLuttinger}\tabularnewline
 & $\gamma_{3}=-0.3668$ & \cite{SchoeneLuttinger}\tabularnewline
 & $\eta_{1}=-0.020$ & \cite{SchoeneLuttinger}\tabularnewline 
 & $\eta_{2}=-0.0037$ & \cite{SchoeneLuttinger}\tabularnewline
 & $\eta_{3}=-0.0337$ & \cite{SchoeneLuttinger}\tabularnewline
fourth Luttinger parameter & $\kappa = -0.5$ & \cite{frankmagnetoexcitonscuprousoxide}\tabularnewline
  g-factor of cond. band & $g_{\mathrm{c}}=2.1$ & \cite{ArtyukhinGFactor}\tabularnewline
\hline
\label{tab:Constants}
\end{tabular}
\par\end{centering}

\end{table}

\subsection{Faraday and Voigt configuration, Magneto-Stark-effect}
We consider two different relative orientations of the magnetic field
to the optical axis. In the Faraday configuration, both axes are aligned to be parallel, whereas in the Voigt configuration, they are orthogonal to each other.
Generally, the exciting laser will transfer a finite momentum $\hbar
\boldsymbol{K}$ onto the exciton. This center of mass momentum would
have to be added in the terms for the kinetic energies. Even without a
magnetic field, this leads to quite complicated formulas (cf.\ the
expressions for the Hamiltonian in the supplemental material of
Ref.~\cite{frankjanpolariton}) which are further complicated by the
minimal substitution.
Since the effect of many of the arising terms is presumably negligible due to the smallness of $K$, we simplify the problem and only consider the leading term \cite{frankmagnetoexcitonsbreak,ruder1994atoms}
\begin{equation}
  H_{\mathrm{ms}} = \frac{\hbar e}{M} (\boldsymbol{K} \times \boldsymbol{B})\cdot \boldsymbol{r}
  \label{eq:Magneto-Stark-term}
\end{equation}
in our numerical calculations, which is the well-known motional Stark effect term of the hydrogen atom. This term has the same effect as an external electric field \eqref{eq:MotionalStarkField} perpendicular
to the plane spanned by the wavevector $\boldsymbol{K}$ and the magnetic field vector $\boldsymbol{B}$. Evidently then, the significance of this term depends on the used configuration.

For the Faraday configuration, the effective electrical field~\eqref{eq:MotionalStarkField}
vanishes. A previous investigation of Schweiner \emph{et al.}\ \cite{frankmagnetoexcitonscuprousoxide} was thus conducted under the approximation of vanishing cent{e}r of mass momentum.
They report a complicated splitting pattern where the magnetic field lifts all degeneracies. For a magnetic field oriented along one of the high symmetry axes of the crystal, the symmetry of the exciton is reduced from
$O_{\mathrm{h}}$ to $C_{4\mathrm{h}}$. Still, some selection rules remain, and not all lines become dipole-allowed. Parity remains a good quantum number and since only states with an admixture of P states have nonvanishing oscillator strengths,
only states with odd values of L contribute to the exciton spectrum.

In the Voigt configuration on the other hand, the excitons have a nonvanishing momentum perpendicular to the magnetic field and the Magneto-Stark term has to be included.
For our calculations, we therefore include an electric field, the size of which is given by the wavevector $K_0 = 2.79 \times 10^7 \frac{1}{\mathrm{m}}$ of the incident light and the magnetic field. This value is obtained
by the condition
\begin{equation}
 \frac{\hbar c K_0}{\sqrt{\varepsilon_{b2}}} = E_g - \frac{R_{\mathrm{exc}}}{n^2}
\end{equation}
for $n=5$ and with $\varepsilon_{b2} = 6.46$
\cite{LandoltBornstein1998DielectricConstant} and $R_{\mathrm{exc}} =
86 \, \mathrm{meV}$ \cite{SchoeneLuttinger}, i.e., $\hbar K_0$ is the
momentum of a photon that has the appropriate energy to create an
exciton in the energy range we consider. {Note that, in contrast to Table \ref{tab:Constants}, we here use the dielectric constant in the high frequency limit to describe the refractive index of the incident light.}
Since the total mass $M$ of the exciton is some three orders of magnitudes smaller than for a hydrogen atom, this term will have a significant effect on the spectra, even more so if we consider that the region of high fields is shifted to much lower values for the exciton \cite{frankmagnetoexcitonscuprousoxide}.
The term \eqref{eq:Magneto-Stark-term} breaks the inversion symmetry and parity ceases to be a good quantum number. While in the Faraday configuration only the dipole-allowed exciton states of odd angluar momentum have been important, now also the states of even angluar momentum need to be considered.
Hence, we need to include the terms for the central cell corrections with the Haken potential as given in Refs.~\cite{frankevenexcitonseries,frankjanpolariton} in our treatment to correctly take the coupling to the low lying S states into account.

{In general, polariton effects have to be considered when the center of mass momentum $\boldsymbol{K}$ is nonzero. The experimental results in Refs.~\cite{PhysRevLett.91.107401,doi:10.1002/pssc.200460331,PhysRevB.70.045206} on the other hand show,
that the polariton effects for the $1S$ state are of the order of $10\, \upmu \mathrm{eV}$ and thus small
in comparison with the effects considered in this paper. Furthermore, a recent discussion by Stolz \emph{et al.}\ \cite{1367-2630-20-2-023019} concluded that polariton effects should only be observable in transmission experiments for $n \geq 28$. Hence, we will not include
them in our discussion.}

\section{Numerical approach}
Using the Hamiltonian \eqref{eq:Hamiltonian} with the additional terms for the central cell corrections $H_{\mathrm{CCC}}$ and the Magneto-Stark effect $H_{\mathrm{ms}}$ with a suitable set of basis vectors, the Schr\"odinger equation can be brought into the form of a generalized eigenvalue equation
\begin{equation}
 \boldsymbol{D} \boldsymbol{c} = E \boldsymbol{M} \boldsymbol{c}.
 \label{eq:GeneralizedEigenvalueEquation}
\end{equation}
We choose a basis consisting of Coulomb-Sturmian functions with an appropriate part for the various appearing spins and angular momenta. Due to the broken inversion symmetry, it is not sufficient to include only basis functions of
odd parity as in reference \cite{frankmagnetoexcitonscuprousoxide}. Instead, basis functions of even symmetry have to be included as well.
The resulting equation can then be solved using a suitable LAPACK routine \cite{lapackuserguide3}. For details we refer to
the discussions in Refs.~\cite{frankmagnetoexcitonscuprousoxide,frankevenexcitonseries,PaperMotionalStark}.

\subsection{Oscillator strengths}
The extraction of the dipole oscillator strengths is performed analogously to the calculation for the Faraday configuration \cite{frankmagnetoexcitonscuprousoxide}. For the relative oscillator strengths
we use
\begin{equation}
f_{\mathrm{rel}}\sim\left|\lim_{r\rightarrow0}\frac{\partial}{\partial r}\left\langle \pi_{x,z}\middle|\Psi\left(\boldsymbol{r}\right)\right\rangle\right|^2\label{eq:frel}
\end{equation}
for light linearly polarized in $x$- or $z$-direction. The states $\left| \pi_{x,z} \right\rangle$ are given by
\begin{subequations}
\begin{align}
\left|\pi_x\right\rangle= &\; \frac{i}{\sqrt{2}}\left[\left|2,\,-1\right\rangle_D+\left|2,\,1\right\rangle_D\right],\\
\left|\pi_z\right\rangle= &\; \frac{i}{\sqrt{2}}\left[\left|2,\,-2\right\rangle_D-\left|2,\,2\right\rangle_D\right],
\end{align}
\label{eq:Dxz}%
\end{subequations}
where $\left|F_t,\,M_{F_t}\right\rangle_D$ is an abbreviation \cite{frankmagnetoexcitonscuprousoxide} for
\begin{eqnarray}
& & \left|\left(S_{\mathrm{e}},\,S_{\mathrm{h}}\right)\,S,\,I;\,I+S,\,L;\,F_t,\,M_{F_t}\right\rangle\nonumber\\
& = & \left|\left(1/2,\,1/2\right)\,0,\,1;\,1,\,1;\,F_t,\,M_{F_t}\right\rangle.
\end{eqnarray}
In this state, the electron and hole spin $S_{\mathrm{e}}$ and $S_{\mathrm{h}}$ are coupled to the total spin $S$. This is combined first with the quasispin $I$ and then with the orbital
angular momentum $L$ to obtain the total angular momentum $F_t$. $M_{F_t}$ is the projection onto the axis of quantization.

\section{Experiment}
In the experiment, we investigated the absorption of thin Cu$_2$O crystal slabs. Three different samples with different orientations were available:
In the first sample the [001] direction is normal to the crystal surface, in the other two samples the normal direction corresponds to the [110] and [111] orientation, respectively.
The thicknesses of these samples differed slightly from 30 to $50 \, \upmu \mathrm{m}$  which is, however, of no relevance for the results described below. For application of a magnetic field, the samples were
inserted at a temperature of $1.4\,\mathrm{K}$ in an optical cryostat with a superconducting split coil magnet. Magnetic fields with strengths up to $7\,\mathrm{T}$ could be applied with orientation either parallel
to the optical axis (Faraday configuration) or normal to the optical axis (Voigt configuration).

The absorption was measured using a white light source which was filtered by a double monochromator such that only the range of energies in which the exciton states of interest
are located was covered. A linear polarization of the exciting light, hitting the crystal normal to the slabs, was used. The transmitted light was dispersed by another double
monochromator and detected by a liquid-nitrogen cooled charge coupled device camera, providing a spectral resolution of about $10\,\upmu$eV. Since the spectral width of the studied
exciton resonances is significantly larger than this value, the setup provides sufficient resolution.

\section{Results and discussion\label{sub:field}}
\begin{figure}
\includegraphics[width=1.02\columnwidth]{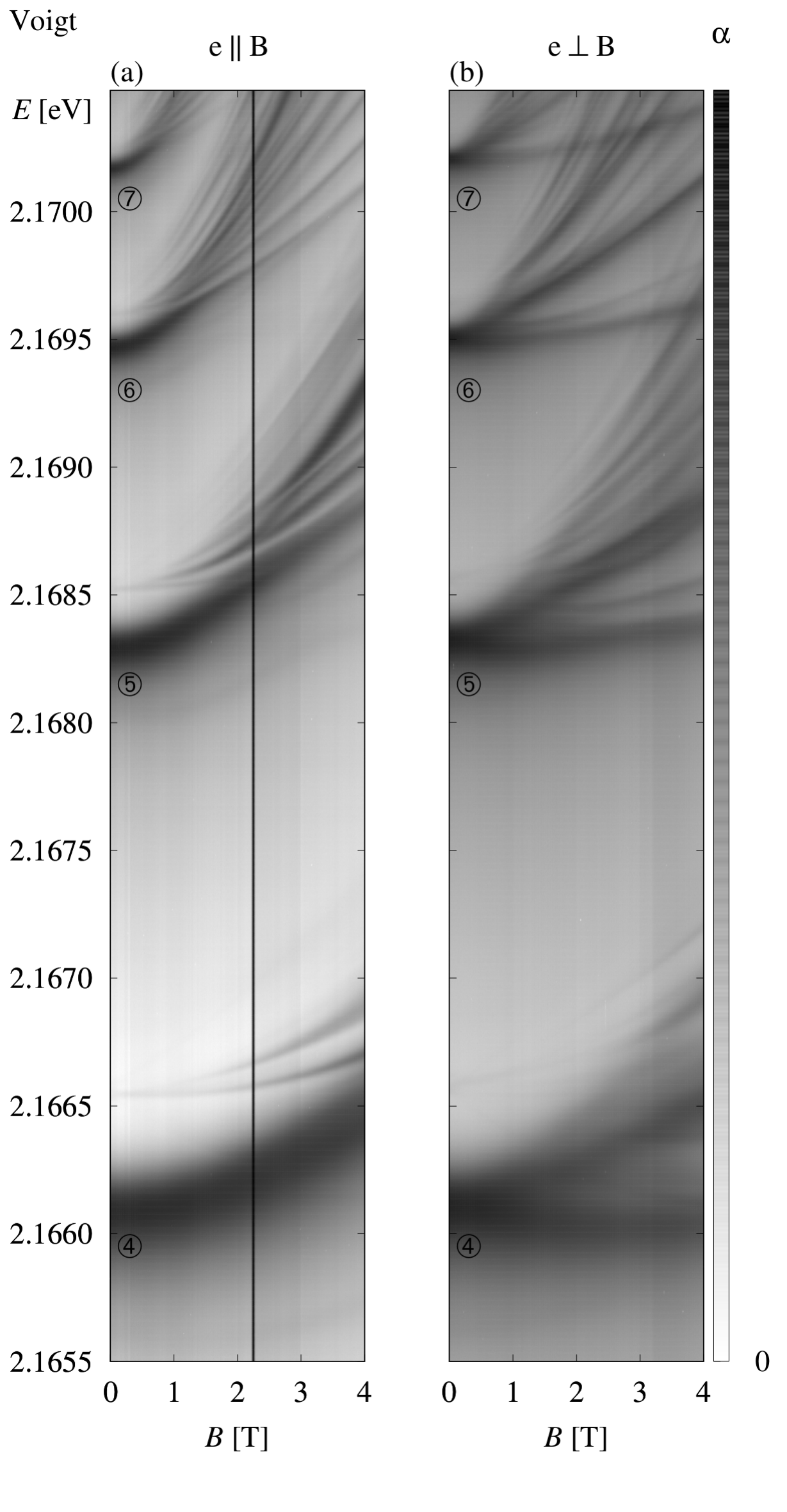}
\caption{{Experimental transmission spectra in arbitrary units for $n=4$ to $n=7$ taken in Voigt configuration for polarization (a) orthogonal [010] and (b) parallel [100] to the magnetic field.}~\label{fig:Voigt_Senkrecht_Parallel_Experiment}}
\end{figure}
\begin{figure}
\includegraphics[width=1.02\columnwidth]{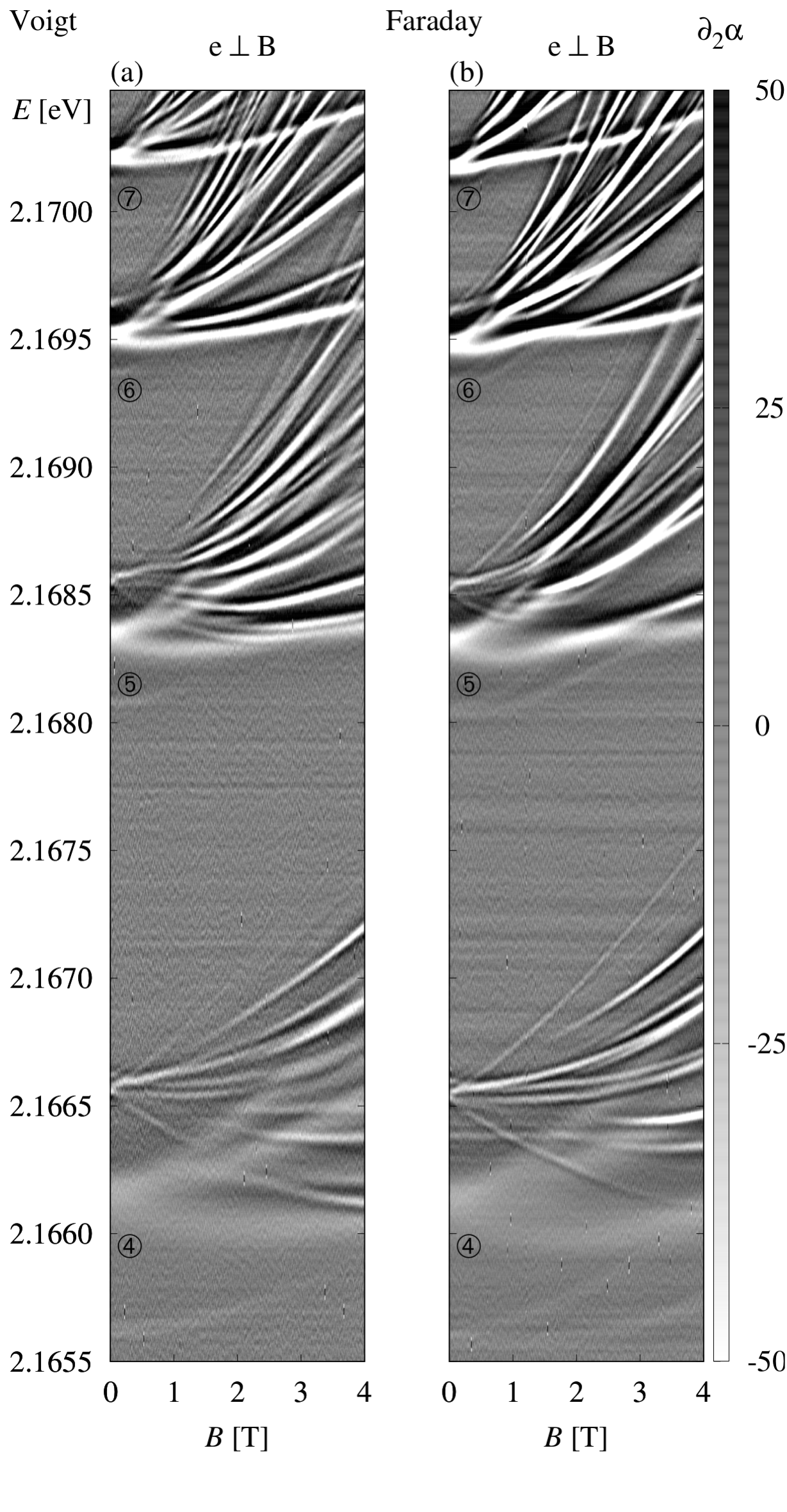}
\caption{{Second derivative of experimental transmission spectra for $n=4$ to $n=7$ taken in (a) Voigt configuration and (b) Faraday configuration with polarization orthogonal [010] to the magnetic field. Data for the Faraday configuration were obtained
by combining $\sigma^+$- and $\sigma^-$-polarized spectra from Ref.~\cite{frankmagnetoexcitonscuprousoxide} in an appropriate linear combination. We use the second derivative for better visibility of weak lines.}~\label{fig:Voigt_Faraday_Experiment}}
\end{figure}
{Figure~\ref{fig:Voigt_Senkrecht_Parallel_Experiment} shows experimental spectra for $n=4$ to $n=7$ in Voigt configuration with polarization orthogonal and parallel to the magnetic field respectively.
The spectra for the two cases show clear differences due to the different selection rules for different polarizations. We will show in Sec.\ \ref{subsec:MagnetoStarkGroupTheory} that all lines in principle become dipole allowed and
can be excited by exactly one of the two polarizations shown here.}

\begin{figure}
\includegraphics[width=1.02\columnwidth]{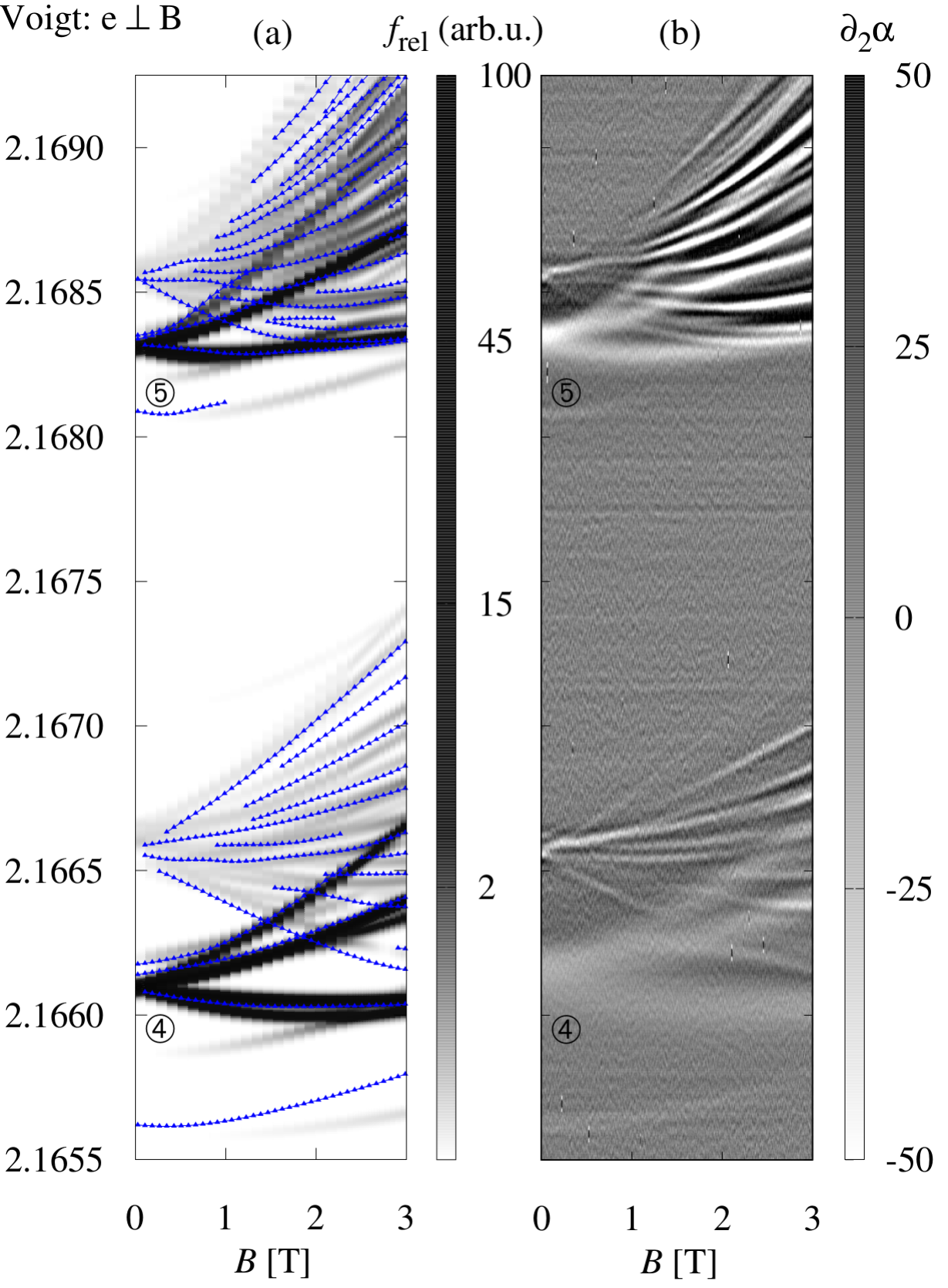}
\caption{{Comparison between numerical and experimental line positions for the Voigt configuration with light polarized orthogonally [010] to the magnetic field. (a) Numerical data in greyscale with read out experimental line positions (blue triangles) and
(b) experimental data using the second derivative to enhance visibility of weak lines. Note that the resolution of the numerical data is not uniform for all field strengths.~\label{fig:VoigtXTheoExpconv}} }
\end{figure}

\begin{figure}
\includegraphics[width=1.02\columnwidth]{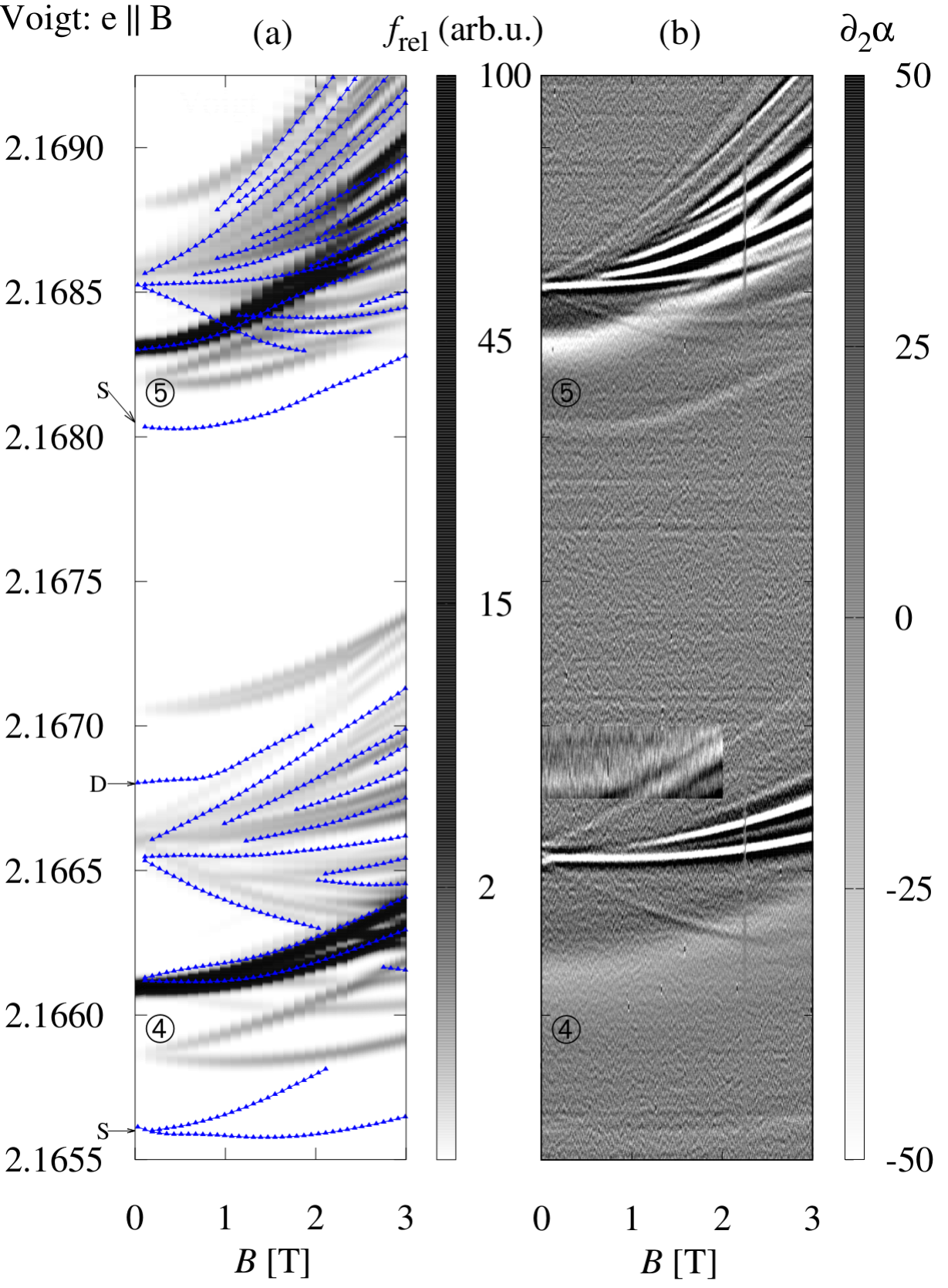}
\caption{{Comparison between numerical and experimental line positions for the Voigt configuration with light polarized parallely [100] to the magnetic field. (a) Numerical data in greyscale with read out experimental line positions (blue triangles) and
(b) experimental data using the second derivative to enhance visibility of weak lines. We increase the visibility of the experimental 4D line by using a different filter width and higher contrast.
Note that the resolution of the numerical data is not uniform for all field strengths.~\label{fig:VoigtZTheoExpconv}} }
\end{figure}

{For the comparison between the Faraday and Voigt configuration we show in Fig.~\ref{fig:Voigt_Faraday_Experiment} experimental spectra taken (a) in Voigt configuration and (b) in Faraday configuration with polarization orthogonal
to the magnetic field respectively.
The polarizations are chosen in such a way that the same selection rules would apply to both spectra in Fig.~\ref{fig:Voigt_Faraday_Experiment} without the Magneto-Stark field. Thus, the differences between them must be due to the different geometries.
S-lines are visible for both configurations. This can be attributed to quadrupole-allowed transitions in the case of the Faraday configuration \cite{frankmagnetoexcitonscuprousoxide}. For the Voigt configuration, these lines
quickly fade away. This is a sign that the additional mixing from the electric field transfers quadrupole oscillator strength away from the S excitons. This effect is not reproduced in the numerical spectra since we only extracted dipole oscillator strengths.
In general, the effective electric field lifts selection rules, revealing additional
lines not visible in the Faraday configuration. This can for example clearly be seen for the $n=5$ states.
}

%{Since the symmetry breaking due to the electric field gets stronger with higher magnetic field strengths, we expect to see S- and D-lines becoming more pronounced with stronger magnetic fields. This behaviour
%differentiates the Magneto-Stark effect from visibility due to quadrupole transitions. Additionally, the effective electric field allows mixing between even and odd states leading to avoided crossings in the spectra.
%}

%
\begin{figure}
\includegraphics[width=1.02\columnwidth]{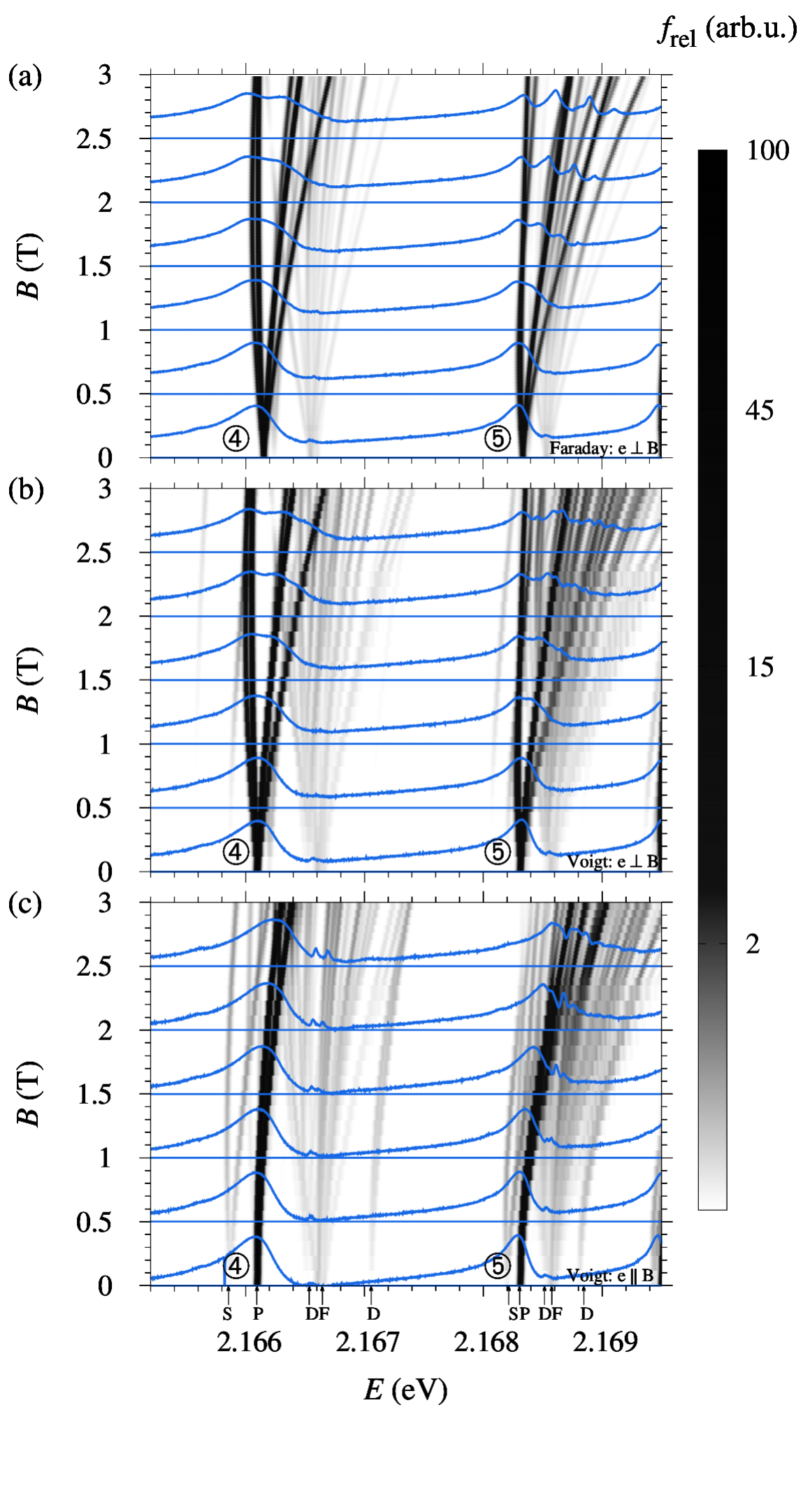}
\caption{Comparison of numerical and experimental spectra of the $n=4$ and $n=5$ excitons in an external magnetic field $\mathbf{B}$~$\parallel$~[100]. (a) {Faraday configuration with light polarized along the [010] direction.
Data were obtained by combining $\sigma^+$- and $\sigma^-$-polarized spectra from Ref.~\cite{frankmagnetoexcitonscuprousoxide} in an appropriate linear combination.} (b) and (c) Voigt configuration with a wavevector aligned with the
[001] direction and the light polarized (b) orthogonally {[010]} and (c) parallely {[100]} to the magnetic field. Numerically calculated relative oscillator strengths are shown in grayscale in arbitrary units.
Experimentally measured absorption coefficients $\alpha$ are superimposed in arbitrary units for a few selected values of B (blue solid lines).
Note that the resolution of the numerical data is not uniform for all field strengths.
We point out the theoretical visibility of S- and D-excitons as marked in (c). {See text for further information.}~\label{fig:VoigtSenkrechtParallelFaraday}}
\end{figure}
{In Figs.~\ref{fig:VoigtXTheoExpconv} and \ref{fig:VoigtZTheoExpconv} we show a comparison between experimental and numerically obtained line positions for $n=4$ and $n=5$.}
{To improve the presentation of areas with many densely lying lines that individually
have very low oscillator strengths, numerical spectra are convoluted using
a Gaussian function with a constant width of $13.6$ $\, \upmu
\mathrm{eV}$.
This value is of the same order of magnitude as the width of the sharpest lines visible in the experiment.}
{While the position of the P- and F-lines is reproduced very well, noticeable disagreement is observed for the S-lines and also the faint 4D-line visible in Fig.~\ref{fig:VoigtZTheoExpconv}.
Since our model is not explicitely constructed on a lattice \cite{Alvermann2018}, we have to include the centrel cell corrections as an approximation into our Hamiltonian. As the centrel cell corrections influence the even parity states
much more strongly than the odd parity states, the error involved in this is more pronounced for the former than for the latter. A similar effect can also be seen in Ref.~\cite{frankevenexcitonseries}.}
{To make additional comparision involving the oscillator strengths possible} we {also} present in
Fig.~\ref{fig:VoigtSenkrechtParallelFaraday}
(a) {data with light linearly polarized orthogonally [010] to the magnetic field} in Faraday configuration taken from Ref.~\cite{frankmagnetoexcitonscuprousoxide} and in (b) and (c) spectra in the Voigt
configuration with light polarized orthogonally {[010]} and parallely {[100]} to the magnetic field axis,
respectively, for the principal quantum numbers $n=4$ and $n=5$.
{
%To get a better agreement between experimental and theoretical line positions, the numerical values are shifted to lower energies by $- 60 \, \upmu \mathrm{eV}$.
%This is equvivalent to a lowering of the energy gap $E_{\mathrm{g}}$ and is justified because of the experimental uncertainties involved in determining this value.
The experimental absorption coefficients do not fall to zero far away from the peaks due to phonon background. We lowered the values with a constant shift to counteract this effect.}
Note that we investigate a parameter region where the effects of quantum chaos as discussed in Ref.~\cite{AszmannGUE2016} are not important.

In general, a good agreement between the experimental and numerical data sets is obtained.
In the Voigt configuration in {Figs.~\ref{fig:VoigtXTheoExpconv}, \ref{fig:VoigtZTheoExpconv} and \ref{fig:VoigtSenkrechtParallelFaraday}}, a rich splitting is observed, especially of the F-states of the $n=5$ excitons. We see that light polarized orthogonally to the magnetic field
probes complementary lines to the ones excited by light polarized in the direction of the field, a result that will also follow from our discussion below.

In experiment, we are not able to resolve the multiplicity of lines that the calculations reveal. This is related to the increased linewidth of the individual features arising
from exciton relaxation by radiative decay and phonon scattering that are not included in the model. Still the field dependences
of the main peaks with largest oscillator strength are nicely reproduced as are the broadenings of the multiplets due to level splitting.

\subsection{Influence of the Magneto-Stark effect}
\label{subsec:MagnetoStarkGroupTheory}
In this section we want to discuss the effects of the additional effective electric field on the line spectra. As we will see in the following group theoretical derivations,
the most pronounced effect is a significant increase in the number of dipole-allowed lines due to the decreased symmetry with the electric field. Panels (a) and (b) in Fig.~\ref{fig:VoigtSenkrechtParallelFaraday}
show this quite clearly, especially for the large number of additional F-lines and also G-lines for $n=5$ in the Voigt configuration. {This is most obvious for the theoretical spectra, but can also distinctly be seen in the experiment for $n=5$.}
Note that without the Magneto-Stark effect the same selection rules would apply to the spectra in (a) and (b), but not in (c).
In contrast to the Faraday configuration \cite{frankmagnetoexcitonscuprousoxide}, we can not limit ourselves to the states with
odd values for L, owing to the mixture of the even and odd series in the electric field. We discuss the case of a magnetic field aligned in [001] direction and will disregard the influence of the central cell corrections in this discussion.

We consider the reduction of the irreducible representations $\tilde{D}^{F \pm}$ of the full rotation group in the presence of the crystal as well as the magnetic and effective electric field, where $F = J + L = (I + S_{\mathrm{h}}) + L$ is the angular momentum without the electron spin.
{Here, the quasispin $I$ and hole spin $S_{\mathrm{h}}$ are first coupled to the effective hole spin $J$ and then combined with the orbital angular momentum $L$ to form $F$.} With this information we will be able to deduce the splitting of the lines due to the reduced symmetry \cite{abragam2012electron}.
Additionally we can compare the resulting irreducible representations with those that the dipole operator belongs to. This will tell us which lines are dipole-allowed and which are not. Note that the symmetry of the quasispin $I$ in $O_{\mathrm{h}}$  is given by
$\Gamma^+_5 = \Gamma^+_4 \otimes \Gamma^+_2$ \cite{frankmagnetoexcitonscuprousoxide} and therefore all irreducible representations have to be multiplied by $\Gamma^+_2$ in comparison with the case of an ordinary spin.
Keeping this in mind, we have \cite{koster1963properties}
\begin{subequations}
\begin{align}
L=0:\hspace{1.7cm}\nonumber\\
\tilde{D}^{\frac{1}{2}+}=D^{\frac{1}{2}+}\otimes\Gamma_{2}^{+} &\: =\Gamma_{6}^{+}\otimes\Gamma_{2}^{+}=\Gamma_{7}^{+},\\
\displaybreak[1]
\nonumber \\
L=1:\hspace{1.7cm}\nonumber\\
\tilde{D}^{\frac{1}{2}-}=D^{\frac{1}{2}-}\otimes\Gamma_{2}^{+} &\: =\Gamma_{6}^{-}\otimes\Gamma_{2}^{+}=\Gamma_{7}^{-},\\
\displaybreak[1]
\nonumber \\
\tilde{D}^{\frac{3}{2}-}=D^{\frac{3}{2}-}\otimes\Gamma_{2}^{+} &\: =\Gamma_{8}^{-}\otimes\Gamma_{2}^{+}=\Gamma_{8}^{-},\\
\displaybreak[1]
\nonumber \\
L=2:\hspace{1.7cm}\nonumber\\
\tilde{D}^{\frac{3}{2}+}=D^{\frac{3}{2}+}\otimes\Gamma_{2}^{+} &\: =\Gamma_{8}^{+}\otimes\Gamma_{2}^{+}=\Gamma_{8}^{+},\\
\displaybreak[1]
\nonumber \\
\tilde{D}^{\frac{5}{2}+}=D^{\frac{5}{2}+}\otimes\Gamma_{2}^{+} &\: =\left(\Gamma_{7}^{+}\oplus\Gamma_{8}^{+}\right)\otimes\Gamma_{2}^{+}\nonumber \\
 &\: =\Gamma_{6}^{+}\oplus\Gamma_{8}^{+},\\
\displaybreak[1]
\nonumber \\
L=3:\hspace{1.7cm}\nonumber\\
\tilde{D}^{\frac{5}{2}-}=D^{\frac{5}{2}-}\otimes\Gamma_{2}^{+} &\: =\left(\Gamma_{7}^{-}\oplus\Gamma_{8}^{-}\right)\otimes\Gamma_{2}^{+}\nonumber \\
 &\: =\Gamma_{6}^{-}\oplus\Gamma_{8}^{-},\\
\displaybreak[1]
\nonumber \\
\tilde{D}^{\frac{7}{2}-}=D^{\frac{7}{2}-}\otimes\Gamma_{2}^{+} &\: =\left(\Gamma_{6}^{-}\oplus\Gamma_{7}^{-}\oplus\Gamma_{8}^{-}\right)\otimes\Gamma_{2}^{+}\nonumber \\
 &\: =\Gamma_{7}^{-}\oplus\Gamma_{6}^{-}\oplus\Gamma_{8}^{-},\\
\displaybreak[1]
\nonumber \\
L=4:\hspace{1.7cm}\nonumber\\
\tilde{D}^{\frac{7}{2}+}=D^{\frac{7}{2}+}\otimes\Gamma_{2}^{+} &\: =\left(\Gamma_{6}^{+}\oplus\Gamma_{7}^{+}\oplus\Gamma_{8}^{+}\right)\otimes\Gamma_{2}^{+}\nonumber \\
 &\: =\Gamma_{7}^{+}\oplus\Gamma_{6}^{+}\oplus\Gamma_{8}^{+},\\
\displaybreak[1]
\nonumber \\
\tilde{D}^{\frac{9}{2}+}=D^{\frac{9}{2}+}\otimes\Gamma_{2}^{+} &\: =\left(\Gamma_{6}^{+}\oplus\Gamma_{8}^{+}\oplus\Gamma_{8}^{+}\right)\otimes\Gamma_{2}^{+}\nonumber\\
 &\: = \Gamma_{7}^{+}\oplus\Gamma_{8}^{+}\oplus\Gamma_{8}^{+}.
\end{align}
\end{subequations}

We still need to include the spin of the electron which transforms according to $\Gamma^+_6$. For vanishing magnetic field strengths,
the representations belonging to an irreducible representation without the spin are degenerate. Those will be written in brackets.
The reduction \cite{koster1963properties} will only be specified for even parity, since the odd case only changes the sign. We obtain
\begin{subequations}
\begin{align}
\tilde{D}^{\frac{1}{2}+} \otimes \Gamma^+_6 &= (\Gamma^+_2 \oplus \Gamma^+_5),\\
\displaybreak[1] \nonumber\\
\tilde{D}^{\frac{3}{2}+} \otimes \Gamma^+_6 &= (\Gamma^+_3 \oplus \Gamma^+_4 \oplus \Gamma^+_5),\\
\displaybreak[1] \nonumber\\
\tilde{D}^{\frac{5}{2}+} \otimes \Gamma^+_6 &= (\Gamma^+_1 \oplus \Gamma^+_4) \oplus (\Gamma^+_3 \oplus \Gamma^+_4 \oplus \Gamma^+_5),\\
\displaybreak[1] \nonumber\\
\tilde{D}^{\frac{7}{2}+} \otimes \Gamma^+_6 &= (\Gamma^+_2 \oplus \Gamma^+_5) \oplus (\Gamma^+_1 \oplus \Gamma^+_4)\\
                                            &\phantom{=}\oplus (\Gamma^+_3 \oplus \Gamma^+_4 \oplus \Gamma^+_5),\nonumber\\
\displaybreak[1] \nonumber\\
\tilde{D}^{\frac{9}{2}+} \otimes \Gamma^+_6 &= (\Gamma^+_2 \oplus \Gamma^+_5) \oplus (\Gamma^+_3 \oplus \Gamma^+_4 \oplus \Gamma^+_5)\\
                                            &\phantom{=}\oplus (\Gamma^+_3 \oplus \Gamma^+_4 \oplus \Gamma^+_5).\nonumber
\end{align}
\end{subequations}
$\Gamma^+_1$ and $\Gamma^+_2$ are one-dimensional, $\Gamma^+_3$ is two-dimensional, and $\Gamma^+_4$ and $\Gamma^+_5$ are three-dimensional.
So without the field, we have for example fourfold degenerate S-states
and P-states that are split into one fourfold and one eightfold degenerate line. If the magnetic field is switched on, the electric field becomes nonvanishing too.
The symmetry is reduced from $O_{\mathrm{h}}$ to $C_{\mathrm{S}}$ \cite{koster1963properties}. All representations of $C_{\mathrm{S}}$ are one-dimensional, so all degeneracies will be lifted,
just as in the case with only a magnetic field. But in contrast to the Faraday configuration, the symmetry is lowered even further, leading to
a greater mixture of the states. In fact, all lines become dipole-allowed. To see this, we have to consider the reduction
of the irreducible representations of $O_{\mathrm{h}}$ in $C_{\mathrm{S}}$ \cite{koster1963properties,abragam2012electron}. The relevent expressions are
\begin{subequations}
\begin{align*}
&\Gamma_{1}^{+} \rightarrow \: \Gamma_{1},& &\Gamma_{1}^{-} \rightarrow \: \Gamma_{2},\\
&\Gamma_{2}^{+} \rightarrow \: \Gamma_{1},& &\Gamma_{2}^{-} \rightarrow \: \Gamma_{2},\\
&\Gamma_{3}^{+} \rightarrow \: \Gamma_{1}\oplus\Gamma_{1},& &\Gamma_{3}^{-} \rightarrow \: \Gamma_{2}\oplus\Gamma_{2},\\
&\Gamma_{4}^{+} \rightarrow \: \Gamma_{1}\oplus\Gamma_{2}\oplus\Gamma_{2},& &\Gamma_{4}^{-} \rightarrow \: \Gamma_{2}\oplus\Gamma_{1}\oplus\Gamma_{1},\\
&\Gamma_{5}^{+} \rightarrow \: \Gamma_{1}\oplus\Gamma_{2}\oplus\Gamma_{2},& &\Gamma_{5}^{-} \rightarrow \: \Gamma_{2}\oplus\Gamma_{1}\oplus\Gamma_{1}.
\end{align*}
\end{subequations}
The dipole operator belongs to $\Gamma^-_4$ in $O_{\mathrm{h}}$ \cite{koster1963properties} and its reduction therefore includes all appearing representations.
Thus, all  $4n^2$ lines receive nonvanishing oscillator strength, the only limitation being given by the polarization of the incident light, i.e., a given state can either be excited
by radiation polarized in the $z$-direction ($\Gamma_{2}$) or by radiation polarized in the $x$-$y$-plane ($\Gamma_{1}$).

\section{Summary\label{sec:Summary}}
We extended the previous work by Schweiner \emph{et al.}\
\cite{frankmagnetoexcitonscuprousoxide} on the optical spectra of
magnetoexcitons in cuprous oxide to the Voigt configuration and
showed that the nonvanishing exciton momentum perpendicular to the
magnetic field leads to the appearance of an effective Magneto-Stark
field.
Including the valance band structure and taking into account central
cell corrections as well as the Haken potential allowed us to produce
numerical results in good agreement with experimental absorption spectra.
We observe a significant increase in the number of visible lines in
both our experimental as well as our numerical data as compared to the
Faraday configuration.
Using group theoretical methods, we show that this is related to the
Magneto-Stark field increasing the mixing between states.
While their positions remain relatively unaffected, the mixing of
states leads to finite oscillator strength of, at least in principle,
all lines.

\acknowledgments
This work was supported by Deutsche Forschungsgemeinschaft
(Grants No.\ MA1639/13-1 and No.\ AS459/3-1).

%\bibliography{Literatur}

\end{document}